# Asymptotically consistent prediction of extremes in chaotic systems:1 stationary case

M. LuValle

Rutgers University

In many real world chaotic systems, the interest is typically in determining when the system will behave in an "average" way or in some extreme manner. Flooding and drought, extreme heatwaves, large earthquakes, and large drops in the stock market are examples of the extreme behaviors of interest. For clarity, in this paper we confine ourselves to the case where the chaotic system to be predicted is stationary so theory for asymptotic consistency can be easily illuminated. We will start with a simple case, where the attractor of the chaotic system is of known dimension so the answer is clear from prior work (Sauer et al[1], and Eckmann and Ruelle[2]). Some extension will be made to stationary chaotic system with higher dimension where a number of empirical results (Bradley and Garland[3], and Ye and Sugihara[4]) will be described and a theoretical framework proposed to help explain them.

To begin, we review some of the results of Sauer et al that allow the first synthesis that results in asymptotically consistent predictions of extremes of a chaotic system. We start with Sauer et al.'s [2] theorem 2.7. The precise statement of the result requires a few mathematical definitions. The box dimension of a compact set A in an n dimensional Euclidean space is $box\ dim(A) = \lim_{\varepsilon \to 0}\left(\frac{log(N_\varepsilon)}{-log(\varepsilon)}\right)$ where $N_\varepsilon$ is the number of cubes with side $\varepsilon$ that it takes to cover A. Again following Sauer (ibid) a general delay coordinate map takes the form $F(x) = \left(h_1(x), \ldots, h_1(g^{p_1-1}(x)), \ldots, h_j(x), \ldots, h_j(g^{p_j-1}(x))\right)$, where $h_i(x)$ is the ith coordinate of a measurement of a term on a chaotic attractor, and $g^k(x)$ is k fold application of a diffeomorphism or flow on A. Let $p = p_1 + \ldots + p_j$. Theorem 2.7 [2] now states, given some restrictions on periodic orbits, that if $p > 2\ box\ dim(A)$ then F is both 1-1 on A and an *immersion* (derivatives are 1-1) on every smooth manifold C contained in A.

Consider the following construction: Take a delay coordinate map where we identify the furthest coordinate forward in time as a statistically dependent variable (y) which we want to model. The remainder are p independent variables. By simply:

1) applying nearest neighbor regression on these p variables to predict y (Stone et al[5]),
2) by letting the number of neighbors in the regression go to infinity as the number of observations
3) assuming the conditions for ergodicity [2] ,as the observing time grows and the number of nearest neighbors goes to infinity and
4) ensuring the number of neighbors =o(number of observations)

We can ensure consistent estimate of the new value as the mean of the data . This however is not much help in extrapolating to new extremes with new dependent variables. To accomplish the extrapolation, we apply the immersion property. Instead of taking the mean of the nearest neighbor dependent variables, we calculate a linear regression of the dependent variable on the p independent ones defining the nearest neighbors. In practice care has to be taken to do this in a statistically sensible manner as the variation in the independent variables are small but because of the immersive nature the regression surface is converging to the tangent plane of the mapping, which will provide an extrapolation correct to the 1st order rather than the 0th order. The ergodic nature of chaotic system[2] and topological nature of delay maps [1] thus ensures asymptotically consistent prediction for extreme values for example taking an m nearest neighbor regression using p independent variables with m=o(number of observations) and p > 2*boxdim(A).

Now let us consider the case where boxdim(A) is unknown. The obvious first step is to let p grow as o(m) in the above scenario, but we would like to know that along the way before p> 2*boxdim(A) we can get useful predictions. There is further theory in Sauer (theorem 2.10) that demonstrates very small probability is assigned to the region where non-immersion occurs for boxdim<p<2*boxdim(A). However Garland and Bradley have shown that useful predictions can arise even with p ( the number of independent variables)=1 in higher dimensional systems.

What is happening in this case where the dimension of the reconstruction space is much lower than that needed for reconstruction. It helps to start with a full embedding in which the p independent variables are a subset of those needed for a full reconstruction, say p1 such variables. Then if there is a unique SRB measure[6], for the system, the marginal distribution of y, the dependent variable given the p independent variables is the integral across the p1-p variables with respect to the SRB measure. Since the SRB measure is absolutely continuous with respect to lebesgue measure, this preserves differentiability on the collapsed system, so that:

i) If the dynamic system is such that E(Y|x) is a nontrivial function then the resulting nearest neighbor model is asymptotically predictive through the ergodic nature of the attractor

ii) AND if the dynamic system admits a unique SRB measure then E(Y|x) has a tangent plane that the regression on nearest neighbors will converge to and provide prediction.

In this case, if p is not a full embedding, than rather than predicting extremes directly, the distribution of residuals around the plane provides a basis for an asymptotically consistent probabilistic prediction of the extremes.

It turns out we can bound the predictability of the tangent planes a little bit based on a further result of Sauer et al[1], and the some extension [].7

Multiview embedding [4] has shown that using several p variable regressions can result in better accuracy. It can be argued (LuValle [7]) that disjoint sampling is in fact a more efficient way to converge to accurate prediction when using Multiview embedding.

To develop this we turn to Sauer et al, to theorem 2.10[1]. One more definition is required for stating theorem 2.10 [1]. The $\delta$ -distant self intersection set is defined as $\Sigma(F,\delta) = \{x \in A: F(x) = F(y) \text{ for some } y \in A, |x - y| > \delta\}$. Then theorem 2.10 states that if A is compact in $R^k$ with $box\ dim(A) = d$, then if $p \leq 2d$, $\Sigma(F,D)$ has lower box counting dimension at most $2d - p$, and $F$ is an immersion on each compact subset C of an m-manifold contained in A except on a subset of C of dimension at most 2m-n-1.

Now define $F_1, F_2$ to be strictly distinct if the values of their respective $h_i$ do not overlap. Then:

**Corollary 1:** For almost all $F_1, F_2$ strictly distinct with dimensions $p_1, p_2$, if $d < p_1 < p_2 \leq 2d$, then $\Sigma(F_1, \delta) \cap \Sigma(F_2, \delta) = \varphi$.

PROOF: Suppose there is a set of $F_i, F_j$ strictly distinct, with positive probability such that $\varphi \neq \Sigma(F_i, \delta) \cap \Sigma(F_j, \delta)$, for $i, j$. Then it would be possible to construct a set of product maps $(F_i, F_j)$ with positive probability for which $\varphi \neq \Sigma\left((F_i, F_j), \delta\right)$, even though the dimension of the product map is $p^{i,j} = p + p_j > 2d$, where $p^{i,j}$ is the dimension of $(F_i, F_j)$ contradicting theorem 2.7 of Sauer (ibid).

Note that if $F_i, F_j$ are distinct, but not strictly distinct, the outcome then depends on the overlap. If the number of overlapping coordinates is such the $p^{i,j} > 2d$ the result holds. If $p^{i,j} \leq 2d$, the null set may be common, but cannot be larger than $2d - p^{i,j}$.

Suppose we wish to model $x$, a term in $F_i, i = 1,2$ and suppose that we have a consistent estimator of $E(x|F_i)$ (e.g. nearest neighbor estimates are consistent (Stone [5])). Suppose $d < p_i < 2d$. Let $\Sigma_i = \Sigma(F_i, \delta), i = 1,2$. Then we can decompose the attractor into 4 mutually exclusive sets, $S_1 = \{\Sigma_1 \cap \Sigma_2\}, S_2 = \{\Sigma_1 \cap \bar{\Sigma}_2\}, S_3 = \{\bar{\Sigma}_1 \cap \Sigma_2\}, S_4 = \{(\bar{\Sigma}_1 \cap \bar{\Sigma}_2)\}$. Suppose we use our hypothetical consistent estimator on each $F_i$ and look at the pairs of prediction based on each.

Exhibit 1
$$E_4\left((E(x|F_1), E(x|F_2))\right) = (x, x)$$
$$E_3\left(\left((E(x|F_1), E(x|F_2))\right)\right) = (x, E_3(x, F_2))$$
$$E_2\left((E(x|F_1), E(x|F_2))\right) = (E_2(x|F_1), x)$$
$$E_1\left(E(x|F_1), E(x|F_2)\right) = (E_1(x|F_1), E_1(x|F_2))$$

So the true value shows up in regions $S_2, S_3, S_4$. Taking advantage of corollary 1, making $F_1$, and $F_2$ strictly distinct, $S_1 = \varphi$ so the true value of the attractor is at least one of the predictions.

The above disjoint system works only for boxdim(A)<p<2*boxdim(A). Now consider arbitrary p, and consider the expected system given $F_1$, and $F_2$. The dimension of the expected system over this reduced mapping is now a maximum of 2p, just a little too large to recursively extend the theorem. So instead consider the conditional expectation over the 2p-1 variables that best approximates the conditional expectation over the 2p variables we have. Assume the conditional system inherits the properties relevant to Sauer et al's theorem 2.10 with respect to this best approximation. The conditional expectation at any given point x, for either the regression based on $F_1$ or that based on $F_2$ must converge to a predictor at least as good a predictor of the original attractor as that best approximation over the 2p-1 variables. A similar statement may be made to approximations based on p/2-1 variables based on Sauers theorem 2. 7

As the p variables being used for prediction varies the system being approximated varies through conditional expectations under the system. Each set p will also have an associated set of residuals in the the local neighborhood of the history being predicted. Taken together, these provide a probabilistic basis for predicting the extremes. Each residual distribution paired with predictions based on its corresponding set p, will provide a probabilistic prediction of the extreme. If we start with K potential variables, and build multiple random partitions of K into p sets of variables, it is possible to create a predictive distribution of chaotic responses using purely least squares fitting of the nearest neighbor data.

In the summer of 2014 a distribution of precipitation predictions was constructed for a region containing a number of weather stations in the tristate area (New Jersey, New York, and Connecticut). Predictions were made for each weather station using each of a number of delay maps. The distribution was corrected at the beginning of August for the amount of rain that had already fallen in June and July, and the predictive distribution based on equally weighted least squares predictions for August is shown below along with the observed rainfall for august in red. The extreme red line is for Islip, New York where 14 inches of that are the result of one 24 hour period. The numbers in the smooth density sections are conditional probabilities assuming each set of p variables is equally likely to provide a useful prediction.

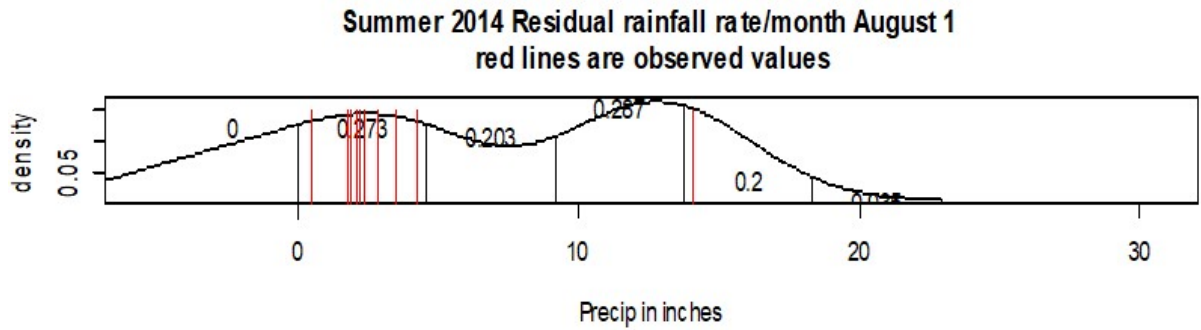

In this case the predictive distribution was not constructed from a history of the process, nor from nearest neighbor linear regression modeling. Instead it was constructed from regression models built around close geographical regions over 100s of years of climate models at constant greenhouse gas concentration. Then subsets of the models where post selected for real data using an evolutionary algorithm (LuValle [8],[9]) to predict precipitation over earlier epochs of real time. While it is not exactly the method described in this paper it illustrates that the approach may be (somewhat) informative and illustrates a simple plot to enable predictive. One of the problems here is the data being plotted is highly correlated, so the predictive density may be reasonable but it requires special evaluation.

[1] Sauer, T., Yoreck, J. and Casdagli, M. "Embedology", *Journal of Statistical Physics*, **65,** 579-616 (1991)
[2] Eckmann, J.P., and Ruelle, D. Ergodic Theory of chaos and strange attractors, *Reviews of modern Physics,* **57,** 617-656, (1985)
[3] Garland J. and Bradley E., "Prediction in Projection", Chaos **25**, 123108 (2015); doi: 10.1063/1.4936242
[4] Ye, H., and Sugihara G.,(2016), "Information leverage in interconnected ecosystems, overcoming the curse of dimensionality", Science, Vol 353, issue 6502, 922-925
[5] C. J. Stone, Ann. Stat. 5, 595–620 (1977).
[6] Young LS, "What are SRB Measures and which dynamical systems have them", Journal of Statistical Physics, vol 108, #516,733-754
[7] LuValle, "A simple statistical approach to prediction in open high dimensional chaotic systems", arXiv, stat, 1902.04727
[8] LuValle, M.J, (2016), "Predicting climate data using climate attractors derived from a global climate model", US Patent 9,262,723
[9] LuValle M. J. (2019), "Statistical Prediction Functions for Natural Chaotic Systems and computer model thereof". US Patent 10,234,595